\newcommand{\be}{\begin{equation}}
\newcommand{\ee}{\end{equation}}
\newcommand{\bea}{\begin{eqnarray}}
\newcommand{\eea}{\end{eqnarray}}
\renewcommand{\l}{\left}
\newcommand{\r}{\right}

\newcommand{\lab}[1]{\label{#1}}
\newcommand{\bi}[1]{\bibitem{#1}}
\newcommand{\da}{\,da}

\newcommand{\dK}{\,d^{3}{\bf k}}
\newcommand{\dKpi}{\frac{d^{3}{\bf k}}{(2\pi)^\frac{3}{2}}}
\newcommand{\ds}{\,ds}
\newcommand{\dt}{\,dt}
\newcommand{\dx}{\,dx}
\newcommand{\DX}{\,{\bf dx}}
\newcommand{\dX}{\,d^{3}{\bf x}}
\newcommand{\Dx}{\,d^{4}x}
\newcommand{\deta}{\,d\eta}
\newcommand{\dltphi}{\,\delta\phi}

\newcommand{\Dltfn}[1]{\,\delta^{3}({\bf #1})}

\newcommand{\phdot}{\dot{\phi}}
\newcommand{\phddot}{\ddot{\phi}}
\newcommand{\half}{\frac{1}{2}}
\newcommand{\thtw}{\frac{3}{2}}
\newcommand{\thir}{\frac{1}{3}}
\newcommand{\twth}{\frac{2}{3}}

\newcommand{\inv}[1]{\frac{1}{#1}}
\newcommand{\smhalf}{{\textstyle \frac{1}{2}}}
\newcommand{\smthtw}{{\textstyle \frac{3}{2}}}
\newcommand{\smthir}{{\textstyle \frac{1}{3}}}
\newcommand{\smtwth}{{\textstyle \frac{2}{3}}}

\newcommand{\smfrac}[2]{{\textstyle \frac{#1}{#2}}}
\newcommand{\sminv}[1]{{\textstyle \frac{1}{#1}}}
\newcommand{\smsqrt}[1]{{\textstyle \sqrt{#1}}}
\newcommand{\aleq}{\mathrel{\rlap{\lower4pt\hbox{\hskip1pt$\sim$}}
                   \raise1pt\hbox{$<$}}}
\newcommand{\ageq}{\mathrel{\rlap{\lower4pt\hbox{\hskip1pt$\sim$}}
                   \raise1pt\hbox{$>$}}}

\documentstyle[twoside,12pt]{article}
\begin{document}
\title{A more accurate analytic calculation of the spectrum of
       cosmological perturbations produced during inflation}
\author{Ewan D. Stewart\thanks{Supported by a JSPS Postdoctoral
        Fellowship} \\ Department of Physics \\ Kyoto University \\
        Kyoto 606, Japan \and David H. Lyth \\
        School of Physics and Materials \\ University of Lancaster
        \\ Lancaster LA1 4YB, U.K.}
\maketitle
\begin{abstract}
Formulae are derived for the spectra of scalar curvature perturbations
and gravitational waves produced during inflation, special cases of
which include power law inflation, natural inflation in the small angle
approximation and inflation in the slow roll approximation.
\end{abstract}
\vspace*{-85ex}
\hspace*{\fill}{\bf KUNS 1176}\hspace*{3.1em}\\
\hspace*{\fill}{\bf LANCS TH 93 01}\\
\hspace*{\fill}{To be published in}\hspace*{0.9em}\\
\hspace*{\fill}{Physics Letters B}\hspace*{1em}
\thispagestyle{empty}
\setcounter{page}{0}
\newpage
\setcounter{page}{2}

\section{Introduction}
The magnification of vacuum fluctuations in the inflaton field into
large-scale curvature perturbations during inflation
\cite{linde,turner} is the most promising method for producing the seed
inhomogeneities necessary for galaxy formation, and the spectrum of
these inhomogeneities, as well as the spectrum of gravitational waves
produced during inflation, are about the only observational tests of
the properties of the inflaton. It is thus important to calculate these
spectra accurately. In this paper we derive formulae for these spectra,
special cases of which give the exact results for power law inflation
\cite{ppsipl,lm,prpl}, the exact results within the small-angle
approximation for natural inflation \cite{ninfl}, and the results to
second order in the slow roll approximation for inflation in general.
Only the power law results have been given previously
\cite{ppsipl,prpl}. The standard results to first order in the slow
roll approximation are
\be
P^{\half}_{\cal R}(k) = \left. \left(\frac{H^{2}}{2\pi\phdot}\right)
\right|_{aH=k}                                                \lab{pr0}
\ee
for the curvature perturbation spectrum \cite{lyth,linde,turner} and
\be
P^{\half}_{\psi}(k) = \left. \left(\frac{H}{2\pi}\right) \right|_{aH=k}
                                                            \lab{ppsi0}
\ee
for the gravitational wave spectrum \cite{star,turner}.

\section{Notation}
Our units are such that
$c=\hbar=8\pi G=1$. $H$ is the Hubble parameter, $\phi$ is the inflaton
field and a dot denotes the derivative with respect to time $t$. The
background metric is
\be
\ds^{2} = \dt^{2}-a^{2}(t)\DX^{2} = a^{2}(\eta)[\deta^{2}-\DX^{2}]
                                                          \lab{metric0}
\ee
Scalar linear perturbations to this metric can be expressed most
generally as \cite{lpert}
\be
\ds^{2} = a^{2}(\eta) \left\{(1+2A)\deta^{2}-2\partial_{i}B\dx^{i}\deta
-\left[(1+2R)\delta_{ij}+2\partial_{i}\partial_{j}H_{T}\right]
\dx^{i}\dx^{j}\right\}                                    \lab{metric1}
\ee
$\cal R$ is the intrinsic curvature perturbation of comoving
hypersurfaces, and, during inflation, is given by
\be
{\cal R} = R-\frac{H}{\phdot}\dltphi                         \lab{rdef}
\ee
where $\delta\phi$ is the perturbation in the inflaton field. On each
scale ${\cal R}$ is constant well outside the horizon. Its
spectrum is defined by
\be
{\cal R} = \int\dKpi {\cal R}_{\bf k}(\eta)e^{i{\bf k.x}}      \lab{rf}
\ee
\be
\left<{\cal R}_{\bf k}{\cal R}^{*}_{\bf l}\right> =
\frac{2\pi^{2}}{k^3}P_{\cal R}\Dltfn{k-l}                   \lab{prdef}
\ee
Tensor linear perturbations to (\ref{metric0}) can be expressed most
generally as \cite{lpert}
\be
\ds^{2} = a^{2}(\eta) \left[\deta^{2}-(\delta_{ij}+2h_{ij})\dx^{i}
\dx^{j}\right]                                            \lab{metric2}
\ee
The spectrum of gravitational waves is defined by
\be
h_{ij} = \int\dKpi\sum_{\lambda=1}^{2}\psi_{{\bf k},\lambda}(\eta)
e_{ij}({\bf k},\lambda)e^{i{\bf k.x}}                         \lab{hij}
\ee
\be
\left<\psi_{{\bf k},\lambda}\psi_{{\bf l},\lambda}^{*}\right> =
\frac{2\pi^{2}}{k^{3}}P_{\psi}\Dltfn{k-l}                 \lab{ppsidef}
\ee
where $e_{ij}({\bf k},\lambda)$ is a polarization tensor satisfying
\be
e_{ij}=e_{ji} \;\;,\;\; e_{ii}=0 \;\;,\;\; k_{i}e_{ij}=0  \lab{polten1}
\ee
\be
e_{ij}({\bf k},\lambda)e_{ij}^{*}({\bf k},\mu)=\delta_{\lambda\mu}
                                                          \lab{polten2}
\ee
It is also useful to choose
\be
e_{ij}(-{\bf k},\lambda)=e_{ij}^{*}({\bf k},\lambda)      \lab{polten3}
\ee

\section{The Calculation}
The effective action during inflation is assumed to be
\be
S=-\smhalf\int R\smsqrt{-g}\Dx+\int\left[\smhalf(\partial\phi)^{2}
-V(\phi)\right]\smsqrt{-g}\Dx                              \lab{action}
\ee
The action for scalar linear perturbations is then \cite{muk,mukrev}
\be
S=\half\int\left[(u')^{2}-(\partial_{i}u)^{2}+\frac{z''}{z}u^{2}\right]
\deta\dX                                                  \lab{saction}
\ee
where $z=\frac{a\phdot}{H}$ and a prime denotes the derivative with
respect to conformal time $\eta$. $u$ is $a$ times the inflaton field
perturbation on spatially flat hypersurfaces and, from (\ref{rdef}),
during inflation
\be
u=-z{\cal R}                                                 \lab{udef}
\ee
Quantizing
\be
\hat{u}(\eta,{\bf x}) = \int\dKpi\left\{u_{k}(\eta)\hat{a}_{\bf k}
e^{i{\bf k.x}}+u_{k}^{*}(\eta)\hat{a}_{\bf k}^{\dag}e^{-i{\bf k.x}}
\right\}                                                     \lab{uhat}
\ee
\be
\left[\hat{a}_{\bf k},\hat{a}_{\bf l}^{\dag}\right] = \Dltfn{k-l} \;\;
,\;\; \hat{a}_{\bf k}|0>=0 \;\;,\;\; {\rm etc.}              \lab{ahat}
\ee
The equation of motion for $u_{k}$ is
\be
u_{k}''+\left(k^{2}-\frac{z''}{z}\right)u_{k}=0              \lab{ueom}
\ee
and
\be
u_{k} \rightarrow \smfrac{1}{\sqrt{2k}}e^{-ik\eta} \;\;\;\;
{\rm as}\;\;\; aH/k \rightarrow 0                            \lab{ulim}
\ee
corresponding to flat spacetime field theory well inside the horizon.
Also the growing mode for $aH/k\gg1$ is
\be
u_{k} \propto z                                               \lab{ugm}
\ee
Now
\be
\frac{z''}{z} = 2a^{2}H^{2}\left(1+\smthtw\delta+\epsilon
+\smhalf\delta^{2}+\smhalf\epsilon\delta
+\smhalf\smfrac{1}{H}\dot{\epsilon}
+\smhalf\smfrac{1}{H}\dot{\delta}\right)                      \lab{zed}
\ee
where
\be
\epsilon\equiv\frac{-\dot{H}}{H^{2}} \;\;,\;\;
\delta\equiv\frac{\phddot}{H\phdot}                            \lab{ed}
\ee
and
\be
\eta = \int\frac{dt}{a} = \int\frac{da}{a^{2}H} =
\frac{-1}{aH}+\int\frac{\epsilon\da}{a^{2}H}                  \lab{eta}
\ee
Thus if $\epsilon$ and $\delta$ are constant, which we shall assume
here, then (\ref{ueom}) can be solved easily:
\be
\eta = \frac{-1}{aH}\left(\frac{1}{1-\epsilon}\right) \;\;\;\;\;\;
({\rm N.B.}\;\;\epsilon<1\;\Leftrightarrow\;{\rm inflation}) \lab{etac}
\ee
\be
\frac{z''}{z} = \frac{1}{\eta^{2}}\left(\nu^{2}-\frac{1}{4}\right)
\;\;\;\;{\rm where}\;\;\; \nu = \frac{1+\delta+\epsilon}{1-\epsilon}
+\half                                                        \lab{znu}
\ee
\begin{eqnarray}
u_{k} & = & \smhalf\smsqrt{\pi}e^{i(\nu+\half)\frac{\pi}{2}}
(-\eta)^{\half}H^{(1)}_{\nu}(-k\eta)                        \lab{us} \\
      & \rightarrow & e^{i(\nu-\half)\frac{\pi}{2}}2^{\nu-\frac{3}{2}}
\frac{\Gamma(\nu)}{\Gamma(\frac{3}{2})}\inv{\sqrt{2k}}
(-k\eta)^{\half-\nu} \;\;\;{\rm as}\;\; aH/k \rightarrow \infty
                                                            \lab{uslim}
\end{eqnarray}
Now from (\ref{rf}), (\ref{udef}) and (\ref{uhat})
\be
<0|\hat{\cal R}_{\bf k}\hat{\cal R}_{\bf l}^{\dag}|0> =
\frac{1}{z^{2}}\left|u_{k}\right|^{2}\Dltfn{k-l}               \lab{ru}
\ee
Therefore from (\ref{prdef}) and (\ref{uslim})
\begin{eqnarray}
P_{\cal R}^{\half}(k) & = & \sqrt{\frac{k^{3}}{2\pi^{2}}}
\left|\frac{u_{k}}{z}\right|                               \lab{pru} \\
                      & = & 2^{\nu-\frac{3}{2}}
\frac{\Gamma(\nu)}{\Gamma(\frac{3}{2})}(1-\epsilon)^{\nu-\half}
\left.\frac{H^{2}}{2\pi\left|\phdot\right|}\right|_{aH=k}     \lab{prc}
\end{eqnarray}

The calculation for the gravitational wave spectrum is very similar.
The action for tensor linear perturbations is \cite{mukrev}
\begin{eqnarray}
S & = & \smhalf\int a^{2}\left[\left(h'_{ij}\right)^{2}
-\left(\partial_{l}h_{ij}\right)^{2}\right]\deta\dX         \lab{sh} \\
  & = & \half\int\dK\sum_{\lambda=1}^{2}\int\left[\left|
v'_{{\bf k},\lambda}\right|^{2}-\left(k^{2}-\frac{a''}{a}\right)\left|
v_{{\bf k},\lambda}\right|^{2}\right]\deta                     \lab{sv}
\end{eqnarray}
where
\be
v_{{\bf k},\lambda} = a\psi_{{\bf k},\lambda}                \lab{vpsi}
\ee
N.B. $v_{{\bf k},\lambda} = v_{{\bf -k},\lambda}^{*}$ from (\ref{hij})
and (\ref{polten3}).
Quantizing
\be
\hat{v}_{{\bf k},\lambda}(\eta) = v_{k}(\eta)\hat{a}_{{\bf k},\lambda}
+v_{k}^{*}(\eta)\hat{a}_{{\bf -k},\lambda}^{\dag}              \lab{va}
\ee
\be
\left[\hat{a}_{{\bf k},\lambda},\hat{a}_{{\bf l},\sigma}^{\dag}\right]
= \delta_{\lambda\sigma}\Dltfn{k-l} \;\;,\;\;
\hat{a}_{{\bf k},\lambda}|0>=0 \;\;,\;\; {\rm etc.}         \lab{tahat}
\ee
The equation of motion for $v_{k}$ is
\be
v''_{k}+\left(k^{2}-\frac{a''}{a}\right)v_{k}=0              \lab{veom}
\ee
and
\be
v_{k} \rightarrow \smfrac{1}{\sqrt{2k}}e^{-ik\eta} \;\;\;{\rm as}\;\;
aH/k \rightarrow 0                                           \lab{vlim}
\ee
\be
v_{k} \propto a \;\;\;\;{\rm for}\;\;\; aH/k\gg1              \lab{vgm}
\ee
As before assuming $\epsilon$ is constant
\begin{eqnarray}
\frac{a''}{a} & = & 2a^{2}H^{2}(1-\smhalf\epsilon)          \lab{ae} \\
              & = & \frac{1}{\eta^{2}}\left(\mu^{2}-\frac{1}{4}\right)
\;\;\;\;{\rm where}\;\;\; \mu = \frac{1}{1-\epsilon}+\half    \lab{amu}
\end{eqnarray}
Now
\be
<0|\hat{\psi}_{{\bf k},\lambda}\hat{\psi}_{{\bf l},\sigma}^{\dag}|0> =
\frac{1}{a^{2}}\left|v_{k}\right|^{2}\delta_{\lambda\sigma}\Dltfn{k-l}
                                                             \lab{psiv}
\ee
Therefore
\be
P_{\psi}^{\half}(k) = 2^{\mu-\frac{3}{2}}
\frac{\Gamma(\mu)}{\Gamma(\frac{3}{2})}(1-\epsilon)^{\mu-\half}
\left.\frac{H}{2\pi}\right|_{aH=k}                          \lab{ppsic}
\ee

\section{Special Cases}
\subsection{Power law inflation}
In power law inflation
\be
a \propto t^{p}                                               \lab{pl1}
\ee
Therefore from (\ref{ed})
\be
\epsilon=-\delta=\frac{1}{p}={\rm constant}                   \lab{pl2}
\ee
Therefore from (\ref{znu}) and (\ref{amu})
\be
\nu=\mu=\thtw+\frac{1}{p-1}                                   \lab{pl3}
\ee
Therefore from (\ref{prc}) and (\ref{ppsic})
\be
P_{\cal R}^{\half}(k) = \l[2^{\inv{p-1}}
\frac{\Gamma\l(\thtw+\inv{p-1}\r)}{\Gamma(\thtw)}
(1-\sminv{p})^{\frac{p}{p-1}}\r]\sqrt{\frac{p}{2}}\frac{H_{1}}{2\pi}
\l(\frac{k_{1}}{k}\r)^{\inv{p-1}}                            \lab{prpl}
\ee
where $H_{1}=H|_{aH=k_{1}}$, in agreement with \cite{prpl}, and
\be
P_{\psi}^{\half}(k)=\smsqrt{\frac{2}{p}}P_{\cal R}^{\half}(k)\lab{pppl}
\ee
in agreement with \cite{ppsipl}.

\subsection{Natural inflation}
In natural inflation \cite{ninfl} the inflaton potential is
\be
V(\phi)=\Lambda^{4}\l[1+\cos\l(\frac{\phi}{f}\r)\r]           \lab{ni1}
\ee
In the small-angle approximation, ie.\ $\frac{\phi}{f}\ll1$, we have
\be
H\simeq\smsqrt{\twth}\Lambda^{2}                              \lab{ni2}
\ee
\be
\frac{dV}{d\phi}\simeq-\frac{\Lambda^{4}}{f^{2}}\phi          \lab{ni3}
\ee
Therefore
\be
\phi\propto\exp\l[\thtw\l(\sqrt{1+\frac{2}{3f^{2}}}-1\r)Ht\r] \lab{ni4}
\ee
\be
\epsilon\simeq0 \;\;\;\; {\rm and} \;\;\;\;
\delta\simeq\thtw\l(\sqrt{1+\frac{2}{3f^{2}}}-1\r)            \lab{ni5}
\ee
\be
P_{\cal R}^{\half}(k) \simeq \l[2^{\delta}
\frac{\Gamma\l(\thtw+\delta\r)}{\Gamma(\thtw)}\r]
\frac{\Lambda^{2}}{\sqrt{6}\pi\phi_{1}\delta}
\l(\frac{k_{1}}{k}\r)^{\delta}                                \lab{ni6}
\ee
\be
P_{\psi}^{\half}(k) \simeq \frac{\Lambda^{2}}{\sqrt{6}\pi}    \lab{ni7}
\ee
giving a spectral index for the scalar curvature perturbation
\be
n_{\cal R} \simeq 1-2\delta                                   \lab{ni8}
\ee
For $n_{\cal R}=0.7$ \cite{dhlrep}, the approximate spectral index
$n_{\cal R} \simeq 1-1/f^{2}$ given in \cite{ninfl} gives a 2\% error
in $f$, which, when combined with using (\ref{pr0}) instead of
(\ref{ni6}), leads to a 60\% error in the predicted value of the
Hubble constant during inflation. This large error is mainly due to the
sensitive dependence of $\phi_{1}$ on $\delta$, $\phi_{1} \sim
e^{-60\delta}$. However, for observational errors not to dominate, the
spectral index would have to be measured to an accuracy of $n_{\cal R}
= 0.7 \pm 0.02$. Note that for $n_{\cal R}=0.7$, $\phi_{1} \sim
10^{-4}$ and so the small angle approximation is much better than the
slow roll approximation.

\newcommand{\phdddot}{\stackrel{\cdot\!\!\!\cdot\!\!\!\cdot}{\phi}}
\newcommand{\fourdots}{\cdot\!\!\!\cdot\!\!\!\cdot\!\!\!\cdot}
\newcommand{\phddddot}{\stackrel{\fourdots}{\phi}}

\subsection{Inflation in general}
To obtain the standard results to first order in the slow roll
approximation, (\ref{pr0}) and (\ref{ppsi0}), $\epsilon=
-\frac{d\ln H}{d\ln a}$ and $\delta=\frac{d\ln \phdot}{d\ln a}$ are
neglected. Here we retain $\epsilon$ and $\delta$ but assume that
they are small and neglect terms quadratic in $\epsilon$, $\delta$
and $\frac{\phdddot}{H\phddot}=\frac{d\ln \phddot}{d\ln a}$.\@ Now
\be
\sminv{H}\dot{\epsilon}=2\epsilon(\epsilon+\delta)           \lab{iig1}
\ee
\be
\sminv{H}\dot{\delta} = \delta\l(\frac{\phdddot}{H\phddot}
-\delta+\epsilon\r)                                          \lab{iig2}
\ee
Therefore $\epsilon$ and $\delta$ are approximately constant for small
$\epsilon$, $\delta$ and $\frac{\phdddot}{H\phddot}$, and so we can use
the results of Section 3. Note that $\epsilon$ and $\delta$ only have
to be treated as constant while the mode $k$ is leaving the horizon so
that (\ref{us}) can interpolate between (\ref{ulim}) and (\ref{ugm}),
in the same way that $H$ is treated adiabatically in the standard first
order calculation. Therefore from (\ref{znu})
\be
\nu\simeq\thtw+2\epsilon+\delta                              \lab{iig3}
\ee
and from (\ref{prc}) to lowest order in $\epsilon$ and $\delta$
\be
P_{\cal R}^{\half}(k) \simeq \l[1+\l(2-\ln2-b\r)(2\epsilon+\delta)
-\epsilon\r]\l.\frac{H^{2}}{2\pi\l|\phdot\r|}\r|_{aH=k}      \lab{iig4}
\ee
where $b$ is the Euler-Mascheroni constant and so $2-\ln2-b
\simeq 0.7296$. Similarly
\be
P_{\psi}^{\half}(k) \simeq \l[1-\l(\ln2+b-1\r)\epsilon\r]
\l.\frac{H}{2\pi}\r|_{aH=k}                                  \lab{iig5}
\ee
where $\ln2+b-1 \simeq 0.2704$.
Treating (\ref{iig4}) and (\ref{iig5}) as adiabatic in $\epsilon$ and
$\delta$ then gives the spectral indices
\begin{eqnarray}
n_{\cal R}(k) & = & 1+\frac{d\ln P_{\cal R}}{d\ln k}      \lab{iig6} \\
              & \simeq & 1-4\epsilon-2\delta-2(1+c)\epsilon^{2}
+\smhalf(3-5c)\epsilon\delta \nonumber \\
              &   & -\smhalf(3-c)\delta^{2}+\smhalf(3-c)
\frac{\phdddot}{H\phddot}\delta                              \lab{iig7}
\end{eqnarray}
where $c \equiv 4(\ln2+b)-5 \simeq 0.08145$, and
\begin{eqnarray}
n_{\psi}(k) & = & 1+\frac{d\ln P_{\psi}}{d\ln k}          \lab{iig8} \\
            & \simeq & 1-2\epsilon-(3+c)\epsilon^{2}
-(1+c)\epsilon\delta                                         \lab{iig9}
\end{eqnarray}
If it is now also assumed that $\frac{\phddddot}{H\phdddot}$ is small
then
\be
\epsilon \simeq \half\alpha-\thir\alpha^{2}+\thir\alpha\beta   \lab{ev}
\ee
\be
\delta \simeq \half\alpha-\beta-\twth\alpha^{2}+\frac{4}{3}\alpha\beta
-\thir\beta^{2}-\thir\alpha\gamma                              \lab{dv}
\ee
\be
\frac{\phdddot}{H\phddot}\delta \simeq
\alpha^{2}-\frac{5}{2}\alpha\beta+\beta^{2}+\alpha\gamma      \lab{ddv}
\ee
where
\be
\alpha \equiv \l(\frac{V'}{V}\r)^{2} \;\;,\;\;
\beta \equiv \frac{V''}{V} \;\;{\rm and}\;\;
\gamma \equiv \frac{V'''}{V'}                                \lab{abgv}
\ee
and so
\begin{eqnarray}
n_{\cal R}(k) & \simeq & 1-3\alpha+2\beta+(\smfrac{11}{3}-\smthtw c)
\alpha^{2} \nonumber \\
              &        & -(7-2c)\alpha\beta+\smtwth\beta^{2}
+\smhalf(\smfrac{13}{3}-c)\alpha\gamma                        \lab{nrv}
\end{eqnarray}
and
\be
n_{\psi}(k) \simeq 1-\alpha-(\smthir+\smhalf c)\alpha^{2}
-\smhalf(\smthir-c)\alpha\beta                                \lab{npv}
\ee

\section{Conclusions}
We have derived corrections to the standard slow roll results for the
spectra of scalar curvature perturbations and gravitational waves
produced during inflation. These quantify the errors in the standard
results. In general they are small but, for example, for natural
inflation with a spectral index $n_{\cal R}=0.7$ \cite{dhlrep} the
results of \cite{ninfl} would predict a value of the Hubble constant
during inflation 60\% too high. However, this error would only become
significant compared to the observational errors if the spectral index
were measured to an accuracy of $n_{\cal R}=0.7 \pm 0.02$.
\\[3ex]
\underline{Acknowledgements}: This work was supported in part by
Monbusho Grant-in-Aid for Encouragement of Young Scientists, No.\
92062.

\end{document}